\documentclass[11pt,a4paper]{article}
\usepackage{times}
\usepackage{latexsym}
\usepackage{csquotes}
\usepackage{graphicx}
\usepackage{avm}    
\usepackage{linguex}

\usepackage{url}

\title{Referring to the recently seen: reference and perceptual memory in situated dialog}
\author{John D. Kelleher\\
  ADAPT Research Centre\\
  ICE Research Institute \\
  Technological University Dublin\\
  {\tt john.d.kelleher@dit.ie} \\\and
  Simon Dobnik \\
  CLASP and FLOV \\
  University of Gotenburg, Sweden \\
  {\tt simon.dobnik@gu.se} \\}

\date{}

\begin{document}
\maketitle
\begin{abstract}
From theoretical linguistic and cognitive perspectives, situated dialog systems are interesting as they provide ideal test-beds for investigating the interaction between language and perception. At the same time there are a growing number of practical applications, for example robotic systems and driver-less cars, where spoken interfaces, capable of situated dialog, promise many advantages.  To date, however much of the work on situated dialog has focused resolving anaphoric or exophoric references. This paper, by contrast, opens up the question of how perceptual memory and linguistic references interact, and the challenges that this poses to computational models of perceptually grounded dialog.
\end{abstract}
\section{Introduction}
Situated language is spoken from a particular point of view within a shared perceptual context \cite{byron:2003}.  In an era where we are witnessing a proliferation of sensors that enable computer systems to \emph{perceive} the world, effective computational models of situated dialog have a growing number of practical applications, consider applications in human-robot interaction in personal assistants, driver-less car interfaces that allow interaction with a passenger in language , and so on. From a more fundamental science perspective,  computational models of situated dialog provide a test-bed for theories of cognition and language, in particular those dealing with the binding/fusion of language and perception in the interactive setting involving human conversational partners and ever-changing environment. 

The history of computational models of situated dialog can be traced back to systems in the 1970's such as SHRDLU   which enabled a user to control a robot arm to move objects around a simple simulated blocks micro-world \cite{winograd73}. Since these early beginnings there has been consistent research on computational models of the interface between language and vision\footnote{Examples of such research spanning the decades include \cite{mckevitt9596:NLV,kelleher:2000,kievit/etal:2001,kelleher:2003,gorniak/roy:2004,kelleher2005contextNLG,kruijffetal:2006,Dobnik:2009dz,tellex2010natural,Sjoo:2011aa,kelleher2011visual,hawes2012towards,dobnik2016model,schutte2017robot,larsson2018grounding}}. A commonality across many of these systems is that they have a primary focus on  grounding (in the sense of  Harnad \cite{Harnad:1990} rather than Clark \cite{clark1991grounding}), the references within a single utterance against the current perceptual context. For example, many of these systems are concerned with grounding spatial references\footnote{Herskovits \cite{herskovits:1986} provides an excellent overview of the challenges posed by spatial language. Many computational models of spatial language are based on the spatial template concept proposed by Logan and Sadler \cite{logan/sadler:1996}; see \cite{gapp:1995,kelleher2005context,costellokelleher:06,Kelleher:2009fk} for examples of spatial template based computational models of the semantics of topological prepositions, and \cite{gapp95:_angle_distan_shape_relat_projec_relat,kelleher/vanGenabith:2006,brenner2007mediating} for computational models of projective prepositions. More recently models based on the concept of an attentional vector sum \cite{regier/carlson:2001,kelleher2011effect}, and the functional geometric framework \cite{Coventry/Garrod:2004,Coventryetal:2005} have been proposed. Another stream of research on spatial language deals with the question of frame of reference modelling and ambiguity 
\cite{radvanskyLogan:1997,kelleher2005cognitive,Dobnik:2014aa,Dobnik:2015aa,schultheis2017mechanisms}}. Some of these systems do maintain a model of the evolving linguistic discourse. However, many of these systems assume a fixed view of the world, and hence the question of how to store perceptions of entities that have not yet been mentioned does not arise as the necessary perceptual information relating to these entities is always present through direct perception of the situation. Consequently, these systems have no perceptual memory, and so cannot handle reference to entities that have been perceived but are no longer visible. Within this context, this paper highlights the challenges posed to computational models of situated dialog in designing models that are capable of resolving references to previously perceived entities. % SD 2019-02-07 17:45:06 +0100: my proposoal would be to treat perceptual model in the same way as dialogue common ground; utterances in dialogue also disappear but new information is matched against them. Furthermore, the matching is interactive between the two modalities. A good model of this was the object store in Companions: linguistic entities were added on the object store and matched with the existing ones. Hence, the dialogue object store always contained a set of resolved entities that the agent believes exist.

Paper structure: Section \ref{sec:refindialog} frames the papers focus on reference, and highlights the role that memory plays in reference within dialog; Section \ref{sec:theoriesOfMemory} overviews some of the main cognitive theories and models of human memory; Section \ref{sec:grounding} expands the focus to include models of reference in situated dialog, including models of data fusion from multiple modalities; Section \ref{sec:perceptualmemory} compares two different approaches to designing computational data structures of perceptual memory (one approach is discrete/local/episodic in nature, the other is an evolving monolithic model of context); Section \ref{sec:conclusions} concludes the paper, where we argue that a blend of these approaches is necessary to do justice to the richness and complexity of situated dialog. 
  
\section{Reference in Dialog}
\label{sec:refindialog}

Referring expression can take a variety of surface forms, including: definite descriptions (``the red chair'', indefinites (``a chair''), pronouns (``it''), demonstratives (``that''). The form of referring expression used by a speaker signals their belief with respect to the status the referent occupies within the hearer's set of beliefs. For example, a pronominal reference signals that the intended referent has a high degree of salience within the hearer's current mental model of the discourse context. 

The term mutual knowledge describes set of things that are taken as shared knowledge by interlocutors, and hence are available as referents within the discourse \cite{mccawley93}. In a situated dialog, an interlocutor may consider an entity to be in the mutual knowledge set if: (i) they consider it to be part of the cultural or biographical knowledge they share with their dialog partner, or (i)) it is in the shared perception of the situation the dialog occurs within. 
% SD 2019-01-31 18:04:17 +0100: What is the relation between shared knowledge and common ground? \cite{Clark:1989tw} The reviews of this paper will be more familiar with Clark.
% JK I think we can switch to common ground relatively easily if that helps

The term \emph{discourse context} (DC) is often used in linguistically focused research on dialog to describe the set of entities available for reference due to the fact that they have previously been mentioned in the dialog: 
\begin{displayquote}
``\emph{The DC has traditionally been thought of as a discourse history, and most computational processes accumulate items into this set only using linguistic events as input}'' \cite[pg. 3]{byron:2003}. 
\end{displayquote}

%This distinction opens up the possibility that the internal structure of these two components may be quite distinct, we will return to this question later in the paper. 
In this paper, we will often distinguish between the mutual knowledge set and the discourse context, where the mutual knowledge set contains the set of entities that are available for reference but which have not been mentioned previously in the discourse, and the discourse context being a record of the entities that have been mentioned previously. Given this distinction between mutual knowledge and the discourse context, the process of resolving a referring expression can be characterized as follows: a referring expression in an utterance introduces a representation into the semantics of that utterance and this representation must be bound to an entity in the mutual knowledge set (in the case of evoking or exphoric references) or in the discourse context (in the case of anaphoric references) for the utterance to be resolved.

% SD 2019-02-07 17:46:18 +0100: Here you are ssuming, when you're differentiating the mutual knowledge from discourse context that both agents have complete and identical access to the scene. In reality this is not the case. Hence, there may not be mutual knowledge. Each agent will maintain a different collection of detected dobjects, just like a slightly different set of beliefs, and hence my a4gument about the similarity of perceptual and linguistic common ground/context.

\begin{figure}
\centerline{
\includegraphics[width=0.9\textwidth]{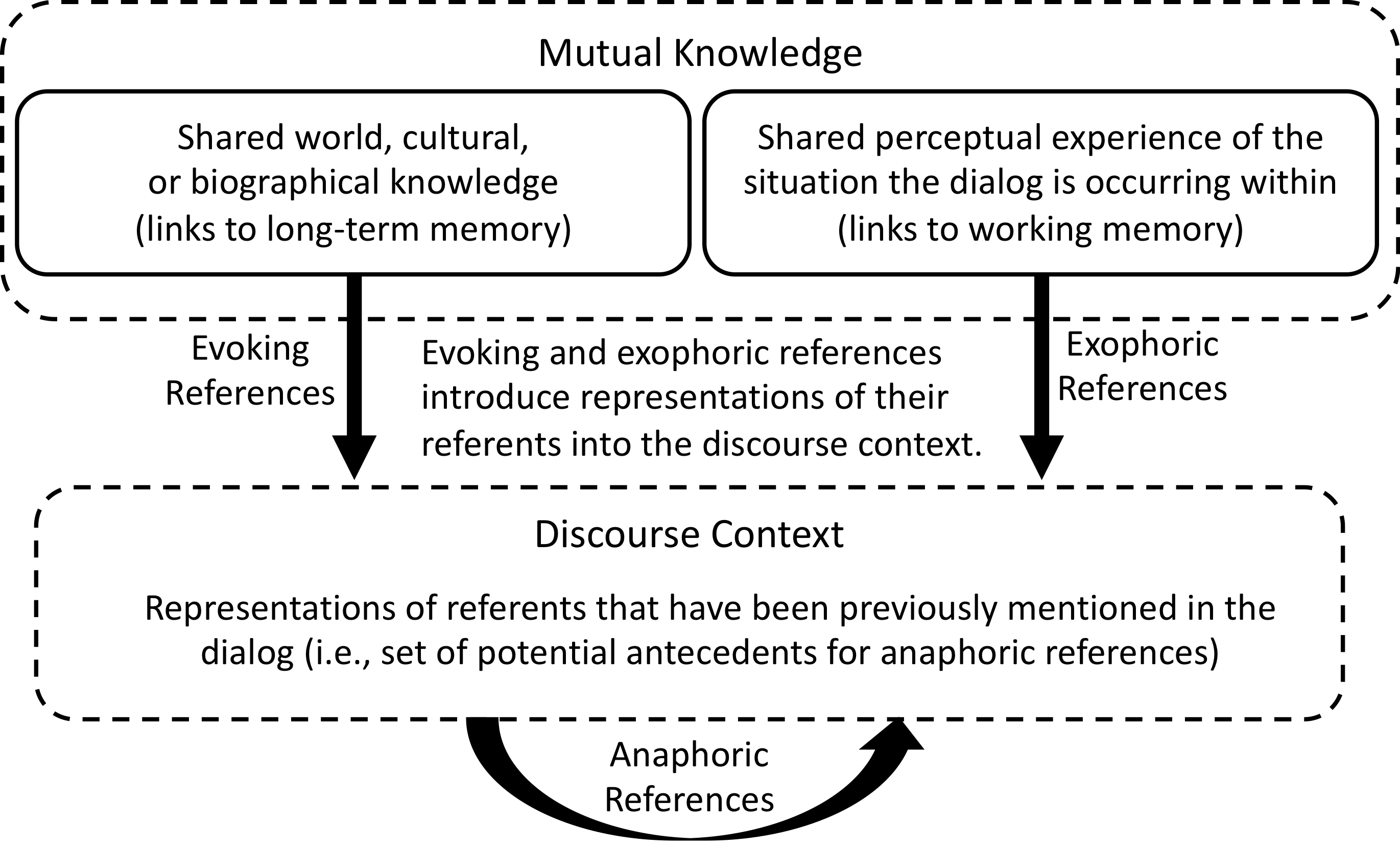}
}
\caption{The relationship between mutual knowledge, the discourse context, and evoking, exophoric, and anaphoric references.} 
\label{fig:MutualDiscourse}
\end{figure}

This process of resolving a referring expression against the mutual knowledge set or the discourse context means that we can distinguish three types of referring expressions based on the information source they draw their referent from (as opposed to their surface) form, namely: \emph{evoking}, \emph{exophoric} and \emph{anaphoric} references. An \emph{evoking} reference refers to an entity that is known to the interpreter through their conceptual knowledge but which has not previously been mentioned in the dialog.  Consequently, the referent of an evoking reference is found in the mutual knowledge set, and the process of resolving this reference introduces a representation of the referent into the discourse context. An \emph{exophoric} reference denotes an entity that is known to the interpreter through their perception of the situation of the dialog but which has not previously been mentioned in the dialog. Similar to an evoking reference, the process of resolving an exophoric reference introduces a representation of the referent into the discourse context. An \emph{anaphoric} reference refers to an entity that has already been introduced mentioned in the dialog and hence a representation of its referent is already in the discourse context. Figure \ref{fig:MutualDiscourse} illustrates the relationships between the data structures and categories of reference described above. 

All of these form of reference draw upon human memory. Mutual knowledge and the maintenance of a discourse context are both 'stored' in memory. Therefore in order for a computational system to be able to resolve exophoric references in must include, and maintain, data structures that represent the the memory component that maintains the mutual knowledge element of shared perceptual experience. To inform the design of this memory data structure in the next section we will review cognitive theories of memory. 

\section{Cognitive Theories of Memory}
\label{sec:theoriesOfMemory}

Cognitive psychology\footnote{See, for example \cite{eysenck2013cognitive}.} distinguishes between a number of different types of memory including:

\begin{description}
	\item[sensory memory] which persists for several hundred milliseconds and is modal specific
	\item[working memory] which persists for up thirty seconds and has limited capacity
	\item[long-term memory] which persists from thirty minutes to the end of an person's lifetime, and has potentially unlimited capacity.
\end{description} 

\begin{figure}
\includegraphics[width=\textwidth]{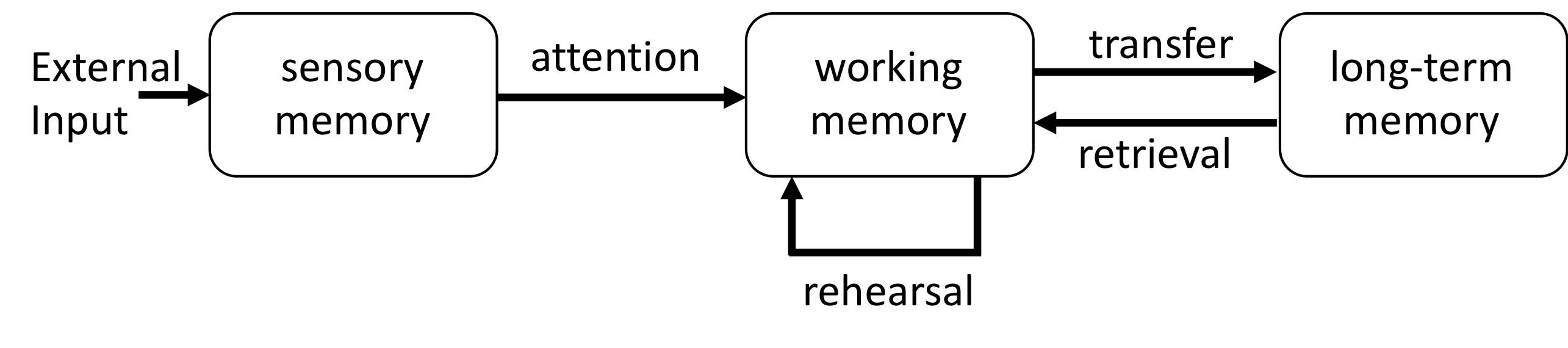}
\caption{Atkinson and Shiffrin's Multi-store Model of Memory} 
\label{fig:atkinsonshiffrin}
\end{figure}

Figure \ref{fig:atkinsonshiffrin} illustrates the Atkinson \& Shiffrin \cite{atkinson1968human} model of how these different types of memory interact . External inputs are initially stored in modality specific sensory memory buffers. There is an attentional filter between these sensory specific memories and working memory. Information that is attended to passes through to working memory, and unattended information is lost. Information in the working memory that is frequently rehearsed is transferred to long-term memory and may be retrieved later. Information is working memory that is not rehearsed is displaced as new information arrives. 

 Evoking references draw on long-term memory and exophoric references draw on working memory\footnote{Exophoric references can also affect the attention filter between sensory memory and working memory, see  \cite{dobnik2016model} for more discussion on this point.}. Furthermore, it is reasonable that the discourse context model should be considered a part of working memory. These observations point to a partial mapping between components of Figure \ref{fig:MutualDiscourse} and Figure \ref{fig:atkinsonshiffrin}. Working memory is where the part of mutual knowledge that is based on perception of the situation and also the discourse context model are stored and maintained; whereas, long-term memory is where the information used to resolve evoking references is stored. The mapping indicates that working memory is at crux of handing exophoric references. 

\begin{figure}
\centerline{
\includegraphics[width=0.65\textwidth]{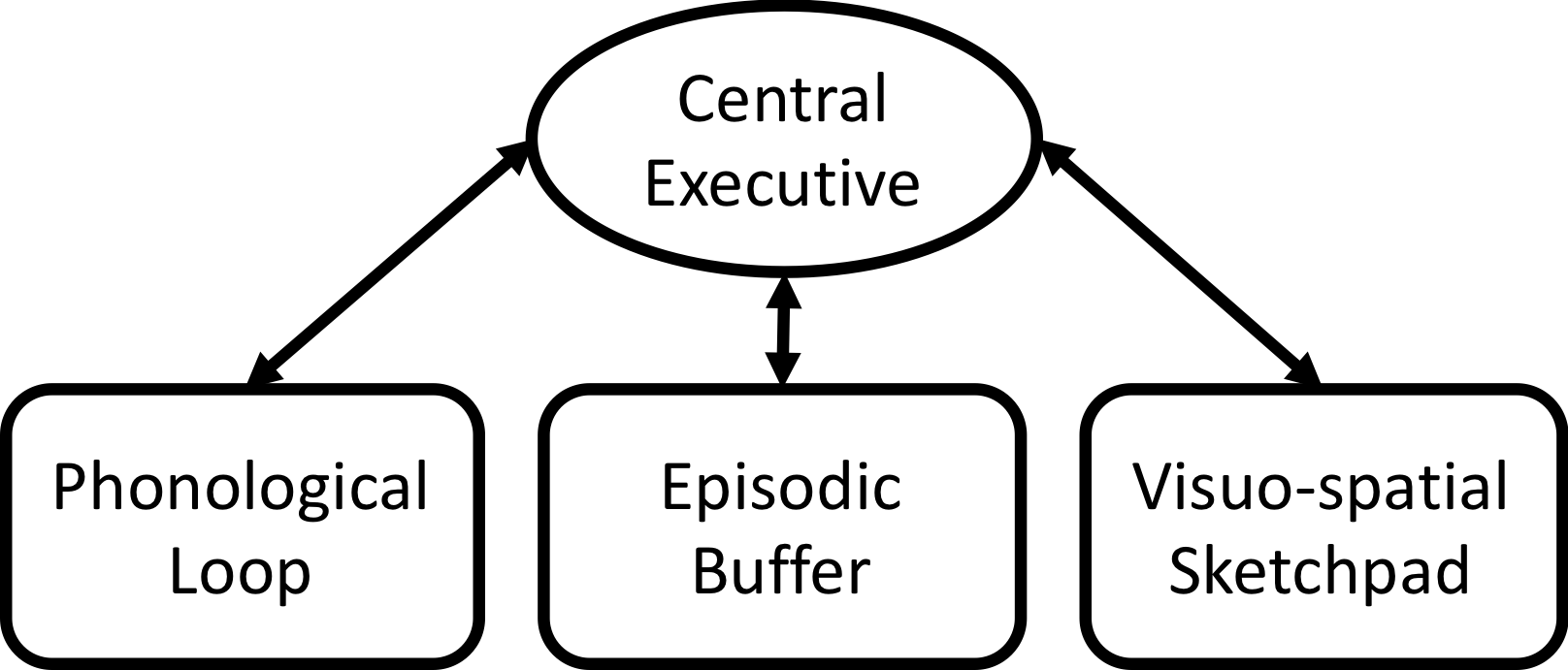}
}
\caption{Baddeley's Model of Working Memory} 
\label{fig:baddeley}
\end{figure}

According to Baddeley \cite{baddeley2002working} working memory has four major systems, see figure \ref{fig:baddeley}, these are:
\begin{description}
	\item[centreal executive] is modality independent and is responsible for supervising the integration of information, directing attention, and coordinating the other systems
	\item[phonological loop] holds speech based information and can maintain this information over short periods by continuous rehearsal
	\item[visual-spatial sketchpad] stores visual and spatial information and can construct visual images and mental maps
	\item[episodic buffer] a limited capacity buffer that temporarily stores and integrates information from the phonological loop and the visuo-spatial sketchpad, and can also link to long-term memory, and perhaps other modules dedicated to smell, taste, and so on. The information sources that the episodic buffer draws upon use different encoding schemes, however the episodic buffer integrates these disparate encodings into a unitary representation of chronologically ordered episodes.
% SD 2019-02-07 17:47:06 +0100: Perhaps this is where the Miller Jonston-Laird 7+-2 takes place.
        \end{description}

\section{Grounding Language in Vision}
\label{sec:grounding}

Grosz \cite{grosz:1977} highlighted that attention processes can affect how references are resolved during a dialog. In particular, Grosz observed the interaction between the shared focus of attention and the use of exophoric definite descriptions. Specifically, if an object is in the mutual focus of attention it can be denoted by means of a definite description even though other entities fulfilling the description are present in the mutual context set. Grosz \& Sidner \cite{grosz/sidner:1986} extended this work and developed a focus stack model of global discourse attentional state. Other models of global discourse structure and processing have since been proposed, for example \cite{hobbs:1985,mann:1987,kempson1988mental,kempson2000dynamic,asher/lascarides:2003,kamp2011discourse}. However, whichever model of global discourse structure is assumed the question of how the focus of attention and reference interact within a local discourse context must also be addressed, and a number of approaches to this question have been proposed, for example \cite{alshawi:1987}, \cite{hajicova:1993}, \cite{lappin/leass:1994}, and \cite{grosz/etal:1995}.\footnote{See
\cite{kruijff-korbayova/hajicova:1997} for a comparison of these
approaches.} However, none of these models explicitly accommodate
multimodal contexts.

Harnad \cite{Harnad:1990} addresses the question of grounding language in perception. More recently, Coradeschi and Saffiotti \cite{coradeschi2003introduction} has addressed this in terms of the symbol anchoring framework, Roy \cite{roy2005semiotic} has proposed semantic schemas, and Kruijff et al. \cite{kruijff/etal:2006} proposed an ontology-based mediation between content in different modalities. Generally, these works focus on exophoric references but assume that the referent is still perceptually available. An interesting, and understudied, category of reference are exophoric references to entities that are not perceptually available at the time of the reference. For example, consider an entity that was seen by two interlocutors just prior to either of them referring to it, but which is no longer visible to either of them, perhaps because they (or it) has changed location. The fact that the entity is no longer accessible through direct perception highlights the need for a memory of perception to be maintained to handle these references, and we will refer to these types of exophoric references as references to perceptual memories. These types of references are interesting for two reasons. First, in general, (as noted above) to date exophoric reference have been studied under the assumption that the referent is still perceptually available to the interlocutors. Second, enabling a computational model to handle exophoric referents to entities that are no longer perceptually available requires the design of a perceptual memory data structure. This perceptual memory data 
structure stores the mutual knowledge information related to the interlocutors shared perceptual experience of the situation (see Section \ref{sec:refindialog}). Furthermore, this perceptual memory data can be understood as part of working memory (see Section \ref{sec:theoriesOfMemory}). 
% SD 2019-02-07 17:47:37 +0100: See my earlier comment about the words also not being available anymore when anaphoric reference is made. Words are also received through perception. The words "red cup", "red cup with a funny top", "Katie's cup", "the cup", that one", "the previous one" all refer to the same object C6.

\section{Perceptual memory}
\label{sec:perceptualmemory}

The design of a perceptual memory data-structure opens up a number of significant research questions, for example: should all entities that are perceived be entered into this data structure or is their a filtering process (e.g. an attentional filter); once and entity enters the perceptual memory is it there indefinitely or can it be removed (forgotten); how does the perceptual memory interact with the linguistic discourse history (are they separate); how is the perceptual memory structured, for example, is it episodic or monolithic, % SD 2019-01-31 18:29:08 +0100: We need to clarify this point. Perhaps this is the time to introduce the discussion in cognitive neuroscience weather maps are global or local/episodic. Meilinger:2018aa and Meilinger:2018ab
does it have a chronological order; and so on. 

There are examples of computational models that can function as perceptual memories in the literature. For example, in Robotics there is a long tradition of research on the problem known as Simultaneous Localisation and Mapping (SLAM), seminal work in this field includes \cite{slam1991}, and \cite{thrun2005probabilistic} provides a more recent introduction and overview of this research. SLAM algorithms integrate sensor information received over a period of time as a robot moves around an environment into a single map representation. Once constructed this map enables a robot to navigate through the environment without colliding with fixed obstacles, such as walls. However, at least in the standard versions of SLAM these maps have no semantic information about what this are, rather the focus is on mapping there are things. So, in some ways, SLAM models can be understood as akin to the visuo-spatial scratchpad in Baddeley's model of working memory. Although undoubtably useful for robot navigation, SLAM models, and the encodings they use, are not designed to facilitate linguistic reference. For this, we need a model that integrates both visuo-spatial information and linguistic information, something akin to the episodic buffer in Baddeley's model. 

% SD 2019-02-07 17:54:21 +0100: I also had another point about SLAM, namely that it builds a single global representation rather than smaller episodic representations. This is also the case with the waypoints which split the global map into more manageable local sub-problems.

\subsection{A Local/Episodic Architecture}
\label{sec:local}

The LIVE system \cite{kelleher/etal:2005}, is a candidate architecture for this episodic buffer module. The LIVE system is designed as a natural language interface to a virtual town, similar in spirit to Winograd's SHRDLU system discussed earlier. A distinctive characteristic of the LIVE system, is that the user was able to move around the environment, and the system had a perceptual memory module that enabled the user to refer to off-screen objects that had been seen recently. The LIVE system uses a false colouring visual salience algorithm to process each frame (visual scene) generated as the user moved through the virtual environment  \cite{kelleher2003false,kelleher/vangenabith:2004}, there are 28 such frames generated per second. This visual salience algorithm identifies each object instance visible in a frame, and associates a normalised visual salience score to each object, based on its size and location within the frame. For each object in a scene the system also retrieves the object type (e.g. house, tree, etc.) and colour information from the scene graph. Consequently, for each frame a list of the visible objects along with their type and colour information and a salience score is created. This frame information is then used to populate a data structure, known as a reference domain. There is a separate reference domain created for each frame. In a sense a reference domain can be understood as a representation of the perceptual information in a frame that is designed to facilitate the grounding of exophoric references. 

\begin{figure}
\centerline{
\includegraphics[width=0.65\textwidth]{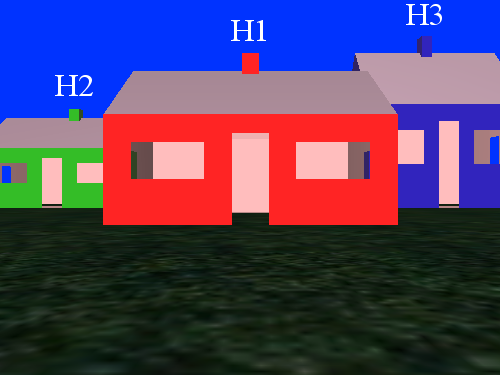}
}
\caption{A frame from the LIVE System. Note: the H1, H2, and H3 labels were added to the image to help readers cross-reference with the reference domain feature structure listed in the paper. } 
\label{fig:liveFrame}
\end{figure}

A reference domain is composed of a number of lists, known as partitions, and the elements of each partition is ordered, in descending order, by their visual salience. The function of these partitions is to predict the different ways a user may refer to an object in the scene. Every reference domain contained a general \emph{object} partition which listed all the objects in the scene ordered by their salience, there was also a partition for each object type in the scene (e.g., if there are trees visible in a frame then the corresponding reference domain would include a tree partition listing all the trees visible ordered by their salience), and for each object colour (e.g., if there are red objects in the scene then there would be a red partition listing all the red objects ordered by colour). The set of potential partitions that could be included in a reference domain is huge, for example there could be a partition for red houses, or green trees, and other combinations of features. In the design of the LIVE system the decision was taken to limit the initial set of partitions to categories that are reasonably likely to be preattentively available, namely, object, type, and colour. Partitions modelling more complex criteria may be created within a reference domain in response to a linguistic utterances, the reasoning being that the act of a referring expression specifying a set of selection restrictions draws attention to the set of objects fulfilling the criteria and therefore creating a partition to explicitly model this set is cognitively plausible at this point. % SD 2019-02-07 18:02:17 +0100: In the CUPS corpus we also observe that objects and regions are created on the fly: "your white ones", "close to you from left to right". This is something that Diedrich Wolter has been working on.
The feature structure below illustrates the reference domain for the frame shown in Figure \ref{fig:liveFrame}.

\begin{center}
\begin{avm}
\[  
p1 & \[ criterion & `object'\\ 
	elements & \[H1,1.0; H3,0.2;H2,0.1\]
	\]\\
p2 & \[ criterion & `house'\\ 
	elements & \[H1,1.0; H3,0.2;H2,0.1\]
	\]\\
p3 & \[ criterion & `red'\\ 
	elements & \[H1,1.0\]
	\]\\
p4 & \[ criterion & `blue'\\ 
	elements & \[H3,0.2\]
	\]\\
p4 & \[ criterion & `green'\\ 
	elements & \[H2,0.1\]
	\]
\]  
\end{avm}
\end{center}

The LIVE system stores these reference domains in a chronologically ordered data structure with a capacity to hold 3,000 reference domains and using a first-in-first-out policy; i.e., when the data structure was full the oldest reference domain was deleted to make space for the new reference domain. This gives the system a perceptual memory of $\frac{3,000}{28}=108$ seconds. 

The LIVE system also maintained a discourse context model. This model is similar in structure to the perceptual memory, it consists of up to 3,000 chronologically ordered reference domain data structures and uses a first-in-first-out policy when the buffer is full. New reference domains are added to this discourse context model as a result of resolving a referring expression. The LIVE system defines different algorithms for resolving referring different forms (i.e. surface forms) of references (i.e, there are separate resolution algorithms for demonstratives, indefinite, definite, pronominal, one anaphora, and other anaphora references). The high-level processing of all of these algorithms is: (i) select a reference domain from either the perceptual memory or the discourse context that contain at least one representation of entity whose features match the selection restrictions in the reference (the selection process also considers the recency and internal structure of the reference domain), (ii) make a copy of the selected reference domain,  (ii) restructure the reference domain (potentially by adding new partitions) to mark the entity selected as the reference, and (iv) add the restructured reference domain to the head of the discourse context list. The restructuring and augmentation of reference domains in response to a referring expression is dependent on the selection restrictions specified in the reference and is designed to facilitate the processing of potential subsequent anaphoric references.

In summary, the LIVE system maintains a separate perceptual memory and discourse context model, although both of these data structure have similar internal structures (chronologically ordered lists of reference domains). The structure of these components is somewhat similar to the episodic Buffer in Baddeley's model: limited capacity, chronologically ordered, and integrating visual perceptual information with semantic information. Furthermore, the similarity in the encodings in the perceptual memory and discourse context model facilitates reference resolution, which entails copying, restructuring, and inserting of a reference domains. Indeed, the approach to resolving a reference taken by the LIVE system can be understood as searching memory for a suitable episodic memory, using this episode as local context within which the reference is resolving, and updating the episode to mark the fact that the reference has occurred. Such a model is capable of handling exophoric references to entities that were recently seen but are no longer on-screen. However, using a reference domain representation of a frame/episode as defining the (local) context for a reference makes it extremely difficult to handle references to refer to two or more entities that never appeared in the same frame. Handling these forms of references requires the system to be able to integrate multiple reference domains, and this is non-trivial; e.g., it is not clear how salience scores from different frames, and hence different times, should be updated during this merger. % SD 2019-02-07 18:14:21 +0100: Weighted by their recency (or age)?

\subsection{A Global/Monolithic Architecture}
\label{sec:global}

An approach to the design of a perceptual memory, that naturally answers the question of how to integrate information from perceptions received across distinct times, is to use an evolving global structure where all referents are stored in a single data structure that is continuously updated to reflect the current state. 

Koller et al. \cite{koller2004put} describes an interface for playing textual computer games, based on description logics and theorem proving.  This model does not have a visual component, instead the information relating to the physical environment of the game world is provided via textual descriptions. However, the game world is never fully observable, and therfore a player's knowledge of the game world increases as they move through the game. The context model proposed in this work is based on Description Logics, and uses a data structure known as the \emph{T-Box} to encode axioms related to concepts and roles (in a sense the ontology of the world), and another data structure known as the \emph{A-Box} to encode the entities (instances of concepts) in the world. Interestingly, the system maintains two A-Box data structures: (i) the game A-Box representing the full current game world state, and (ii) the player's A-Box representing what the player knows about the game world (this A-Box is typically a sub-part of the world A-Box). As the player moves through the game environment and explores new locations new instances are added to the player's A-Box. As a result, the player's A-Box represents a perceptual memory of what they have experienced in the world. Entities in the player's A-Box are marked with the property of \emph{here} when they share the same location as the player (i.e., the player and the entity are both in the same room in the world), \emph{visible} if the entity is deemed to be currently visible to the player, and \emph{accessible}  if the player can currently manipulate the entity. Consequently, the system has the ability to distinguish between entities that are currently visible and entities that are known about but which are not visible. However, the design of the reference resolution algorithms used by the system presupposes that: \emph{players will typically only refer to objects which they can ``see'' in the virtual environment, as modelled by the concept `visible'} \cite[page. 12]{koller2004put}. This assumption allows the resolution algorithm to ignore entities in the world which are known to the player (and, hence are in the player's A-Box) but which are not currently visible when resolving a referring expression. This assumption means that the system cannot handle exophoric references to recently seen entities that are no longer visible, as they are deliberately excluded from the context used to resolve references. It should be noted that this is not a simple assumption to remove from the system. The system has no model of perceptual salience (although it does have a model of linguistic salience). As a result it must use this strict visible/invisible criterion to exclude potential distractor entities (that are in the model of the player's knowledge of the world but which are not currently in the perceptual focus), which if not excluded would make a reference appear unspecified and ambiguous to the system. 

Kelleher \cite{kelleher2006attention} is another natural language interface to a virtual world. It is similar to  \cite{kelleher/etal:2005} in that it uses the same visual salience algorithm to analysis the visual frames the user sees as they navigate through the environment. However, the data structure used to store perceptual memories and discourse structure is very different. This system maintains a single global context model throughout a user's session. Once an entity has been rendered on screen a representation of that entity is introduced in this global context model. There is only ever a single representation of an entity in the global context model. This representation of an entity stores the physical information of the entity (e.g., \emph{type}, \emph{colour}, \emph{size}, and so on) and also stores a visual salience and a linguistic salience score for the entity. The visual salience score is updated after each frame is processed. The visual salience of an entity that is not in the current frame is halved when the frame is processed. As a result the visual salience of an entity drops off once it goes out of (visual) focus (i.e., off-screen), and continues to reduce the longer out of focus it remains. The linguistic salience scoring is based on the assumption that entities that have been mentioned recently are more salient than entities that have not. The particular function used to calculate and update the linguistic salience scores is in the spirit of Centering Theory \cite{grosz/etal:1995} and is similar to the model proposed by \cite{krahmer/theune:2002}. The linguistic salience of an entity is updated after each utterance has been processed. The linguistic salience of any entity not mentioned in an utterance is halved when the utterance is processed. Consequently, similar to the visual salience of an entity, the linguistic salience of an entity drops once it leaves the (linguistic) focus, and continues to drop the longer out of focus it remains. As the above description indicates the representation of an entity in the global context model is a relatively complex feature structure. However, the structure of the global context model itself is minimal, it is simply an unordered set of these entity representations. The fact that the linguistic and visual salience scores are updated based on recency of being visible or mention means that the context model does not need to explicitly model recency. 

% SD 2019-02-07 18:29:10 +0100: The Companions dialogue system is using a similar representation. It starts with linguistic data where all referring expressions are matched to objects in the object store. Hence, co-reference is resolving the reference to an existing object using Kennedy and Bugarev's algorithm -- Anaphora resolution for everyone. The system does not use any perceptual representations as it focuses on converstaions about personal experiences of a user. However, the objetcs in the objetc store are fused at some level with emotion tags which comes from lexical information of words and their sequences and acoustic signal.

Reference resolution in this system is done by calculating an integrated salience score for each entity in the context model, and then selecting the entity with the highest integrated score as the referent. The integrated salience score of an entity is recalculated each time a referring expression is processed. The integrated salience score is calculated in three steps: (i) a reference relative visual salience score is calculated by scaling the standard visual salience score to reflect the fit of the entity with the selection restrictions specified in the expression (e.g., in the simplest case the reference relative visual salience score is set to zero if the entity is of the wrong type to be the referent of the reference); (ii) a reference relative linguistic salience score is calculated in a similar way to the reference relative visual salience score; and (iii) the integrated salience score is then calculated using a weighted sum of the reference relative visual and linguistic salience scores, where the weighting is dependent on the surface form of the referring expression (e.g., for pronominal references the system weights linguistic salience more then visual salience). 

The fact that this monolithic global context model does not encode an episodic (frame based)  structure means that the integration of information from different scenes is straight forward. As a result, this system can handle references to entities that do not appear on screen together. However, this flexibility is at a cost. The loss of the episodic chronological order means that a system using this context model would not be able to handle exophoric references based on chronology (such as \emph{the first blue house we saw}), or co-occurrence within a local temporal context (such as \emph{the car that was in front of the house when the man fell}).

\section{Discussion}
\label{sec:conclusions}

The two approaches to perceptual memory described in Sections \ref{sec:local} and \ref{sec:global} can be understood as exemplars at opposing ends of a spectrum of design choices: one focuses on identifying a local context and resolving the reference within that context, and the other focuses on creating and continuously evolving a global context model. These approaches have complementary strengths and weaknesses. Consequently, it is likely that a blend of these approaches is necessary. This is not surprising as there are many examples in language processing\footnote{Switching between local and global representations, similar to the challenge of modelling long-distance dependencies in sequential data \cite{mahalunkar2018using}} where there is a need to be able to switch from a local focus to a global perspective, and back again, as the context requires.

\bibliography{PerceptualMemories} 
\bibliographystyle{plain}

\end{document}